\documentclass[pre,preprint,doublespace,showpacs,floatfix,superscriptaddress]{revtex4}
 \usepackage{graphicx}
 \usepackage{amsfonts,amssymb}
\usepackage{bm}
 \newcommand{\be}{\begin{equation}}
 \newcommand{\ee}{\end{equation}}
 \newcommand{\bea}{\begin{eqnarray}}
 \newcommand{\eea}{\end{eqnarray}}
\newcommand{\bnn}{\begin{eqnarray*}}
\newcommand{\enn}{\end{eqnarray*}}

\begin{document}

\title{Elasticity of semiflexible polymers in two dimensions}

\author{Ashok Prasad, Yuko Hori, Jan\'{e} Kondev}
 \email{ashok@brandeis.edu, yukoh@brandeis.edu, kondev@brandeis.edu}
\affiliation{Martin Fisher School of Physics, Brandeis University,
Mailstop 057, Waltham, MA 02454-9110, USA}

 \date{\today}

\begin{abstract}
We study theoretically the entropic elasticity of a semi-flexible
polymer, such as DNA, confined to two dimensions. Using the worm-like-chain 
model we obtain an exact analytical expression for the partition function of 
the polymer pulled at one end with a constant force. The force-extension 
relation for the polymer is computed in the long chain limit in terms of 
Mathieu characteristic functions. We also present applications to the interaction between a semi-flexible polymer and a nematic field, and derive the nematic order parameter and average extension of the polymer in a strong field.
\end{abstract}
 \maketitle

\section{Introduction}

Mechanical properties of biomolecules, such as their response to an
applied force, are important for understanding a variety of biological
processes ranging from cell motility to gene regulation.
Detailed mechanical studies of biopolymers using laser tweezers,
magnetic tweezers, atomic force microscopy, and other single molecule
techniques have provided the necessary experimental input for formulating
precise mathematical models of their elasticity.
In the case of double-stranded DNA (dsDNA) the favored theoretical model is the
worm-like chain (WLC), proposed by Kratky and Porod in
1949~\cite{Kratky49}.
Owing to the central role that DNA plays in biology,
the last decade has seen  a significant body of
theoretical work on the WLC model,  as well as its modifications and
extensions.~\cite{Marko95c, Strick00, Cocco02b, Samuel02a,Stepanow04, 
Spakowitz04a,Yan04, Wiggins04,Chakrabarti05}. Considerable experimental evidence
has now accumulated that shows that the worm-like-chain model reproduces the
mechanical behavior of dsDNA in the entropic regime quite 
well~\cite{Bustamante94a, Marko95c, Strick00}.

The effect of confinement on the statistical properties of polymers is
of growing interest to scientists and engineers~\cite{Bae01,Fleer93}. In the
context of the cell, the important observation is that dsDNA is always found
in a state of confinement. Namely, it occupies a volume which is considerably
smaller than the volume it assumes free in solution. Here we examine dsDNA
confined to two-dimensions. While not immediately
of relevance to cell biology, this type of confinement has been studied
recently as an interesting polymer physics problem.
Experiments have probed dynamics and thermodynamics of dsDNA confined to
a mica surface \cite{Rivetti96} and the surface of a lipid bilayer
\cite{Maier99}. More generally, two-dimensional confinement is emerging as
an important experimental method, as it allows for the use of
fluorescence microscopy and atomic force microscopy to
obtain images of individual dsDNA molecules. For example, recent experiments
have made use of dsDNA absorbed on treated mica surfaces
to study the effect of DNA-binding proteins on dsDNA
conformations~\cite{Noort04}.

In this paper  we study  the effect of two-dimensional confinement
on the entropic  elasticity of a worm-like-chain polymer. In particular,
we compute the partition function for the WLC model in the
presence of an applied force when the chain is restricted to two
dimensions. This allows us to derive an exact expression for the
average end-to-end distance of the polymer as a function of applied force,
which is the central result of this paper.

The paper is organized as follows.
In sections \ref{sec:wlcmodel} and \ref{sec:forceext} we introduce the WLC 
model and derive an exact
closed-form expression for the tangent partition function and the
force-extension relation of a WLC  polymer pulled at one end by a constant
force, in the limit of large  polymer length. In section \ref{sec:limits} we 
discuss the asymptotic form of our result in the limit of strong and weak 
force, and
compare it to the approximate force-extension relation in three-dimensions. We 
also give a simple
algebraic approximation to the force-extension relation in two-dimensions. In
section \ref{sec:nematic} we show that a minor transformation converts the 
partition function of section \ref{sec:wlcmodel} into
the partition function of a semi-flexible polymer in a nematic field. We use 
this to calculate the nematic order parameter for the polymer, as well as 
its relative extension in a strong nematic field. 

The results reported here should be of practical use
in stretching experiments where semi-flexible polymers such as DNA are
confined to a surface. On the purely theoretical side, they establish an
interesting connection between the elastic properties of two-dimensional
semi-flexible polymers and Mathieu functions.

\section{The worm-like chain model}
\label{sec:wlcmodel}
The worm-like chain model describes the polymer as an elastic filament
characterized by a rigidity parameter $\kappa$, which has dimensions of length.
For a three-dimensional polymer $\kappa$ is the persistence
length, {\it i.e.}, the decay length associated with the tangent-tangent
correlation function, which for DNA under physiological conditions is about
$50$nm. The energy of bending a small arc of length $s$ of the filament
into a circular segment of radius $R$ is given by $k_B T \kappa s/2R^2$, where
$k_B T$ is the thermal energy (at room temperature $k_B T = 4$pNnm).
If we describe polymer conformations by a smooth continuous curve,
the Hamiltonian of the WLC can be expressed as the integral of this
segmental bending energy over the entire polymer curve. Therefore, when a
polymer is tethered at one end and pulled with a constant force $\bm{F}$ in
the $\hat{\bm{x}}$-direction at the other, the Hamiltonian $H$ can be
written as:
\be H=\int_{0}^{L}ds\left\{
\frac{\kappa k_B T}{2}\left(\frac{d\hat{\bm{t}}}{ds}\right) ^{2} -
\bm{F}\cdot\hat{\bm{t}}\right\} , \label{hamil} \ee
where $\hat{\bm{t}}$ is the
unit tangent vector to the polymer curve, and $L$ is the contour length of the
polymer.
In writing eqn.(\ref{hamil}) we have made use of the standard relation between
the radius of curvature of a curve and the rate of change of its tangent
vector, $|d\hat{\bm{t}}/{ds}| = 1/R$.

Since this polymer is represented by a curve, its partition function is the
sum of Boltzmann weights for all  possible curves, subject to the
constraint of fixed polymer length $L$. In terms of the
tangent vector ${\bm{t}}(s)$ the partition function  can be written as a
path integral,
\be
\mathcal{Z}(\hat{\bm{t_f}},\hat{\bm{t_i}};L)=
\int \mathcal{D}[\hat{\bm{t}}(s)]\exp \left[
-\int_{0}^{L}\!ds\left\{\frac{\kappa}{2}
\left(\frac{d\hat{\bm{t}}}{ds}\right)^{2}
-f {\hat {\bm t}} \cdot {\hat {\bm x}} \right\}\right] ,
\label{wlcpart}
\ee
where $f=\bm{F}\cdot \hat{\bm{x}}/k_B T$ is the reduced force with units of 
inverse length.  The dimensionless combination $\kappa f$ delineates regimes of
high ($\kappa f \gg 1$) and low force ($\kappa f \ll 1$), which we take up in
section~\ref{sec:limits}. Experiments on stretching polymers that involve attaching the two ends of the polymer molecule to beads, constrain 
the first and final tangent vectors, usually making them lie along the 
direction of the force. This is indicated by specifying the initial and final 
tangent vectors, $\hat{\bm{t_i}}$ and $\hat{\bm{t_i}}$, explicitly in 
eqn.(\ref{wlcpart}). In case the ends are free, we will need to integrate over
these tangent vectors to obtain the appropriate partition 
function. This is the case, for example, when the polymer is dissolved in a 
nematic solvent, as we shall see in section \ref{sec:nematic}. 

{}From the partition function, eqn.(\ref{wlcpart}),  we calculate the
average end-to-end
extension of the polymer in the direction of the force, $\langle X
\rangle$. Namely, $\langle X \rangle$ is the conformational average of the
end-to-end
vector $\bm{R} = \int_{0}^{L}ds \hat{\bm{t}}$ projected in the direction of the
applied force, {\it i.e.},
\be
\langle X \rangle = \langle \int_{0}^{L}\!d\,s \,\hat{\bm{t}} \cdot \hat{\bm{x}}
\rangle .
\label{genext}
\ee
{}From eqn.(\ref{wlcpart}) we conclude that
\be
\langle X \rangle = \frac{\partial \ln \mathcal{Z}}{\partial f} .
\label{fext1}
\ee

In two dimensions $\hat{\bm{t}}=(\cos\theta, \sin\theta)$, where 
$\theta(s)$ is the polar angle, and the partition function $\mathcal{Z}$ in
eqn.(\ref{wlcpart}) can be rewritten as,
\be
\mathcal{Z}(\theta_f,\theta_i;L)=\int \mathcal{D}[\theta(s)]\exp \left[
-\int_{0}^{L}ds\left\{\frac{\kappa}{2}\left(\frac{d \theta}{ds}\right)^{2}
-f \cos \theta \right\}\right] ,
\label{wlcpart_theta}
\ee
where $\theta_f$ and $\theta_i$ are now the final and initial tangent angles.
The calculation of $\mathcal{Z}(\theta_f,\theta_i;L)$ is made simple by the
standard connection between the path integral and the Schr\"odinger 
differential equation~\cite{Feynman65}
 \be
\left( \frac{\partial}{\partial s} - \frac{1}{2\kappa}
  \frac{\partial^2}{\partial \theta^2} - f\cos\theta \right) \mathcal{Z}(\theta,
\theta_i; s) = 0  \  \ (s>0),
 \label{se} \ee subject to the initial condition
\be
\mathcal{Z}(\theta,\theta_i;0) = \delta (\theta - \theta_i) . \label{gfbc} \ee
Since eqn.(\ref{se}) is separable in $s$ and $\theta$, we can
express its solutions  in terms of the eigenfunctions $Q(\theta)$
and eigenvalues, $E$, of the corresponding eigenvalue equation,
\be
\left(\frac{1}{2\kappa}\frac{\partial^{2}}{\partial \theta^{2}} +
f\cos\theta \right) Q = - EQ . \label{sse}
\ee

Note that the partition function, $\mathcal{Z}(\theta,\theta_i;s)$, is 
formally equivalent to the path integral for a quantum rotor with moment
of inertia  $\kappa$ in an external field $f$, evolving in imaginary time
$s$ ~\cite{Marko95c}, or, equivalently, to a quantum pendulum in a 
gravitational field~\cite{Condon28}. The correspondence 
is  actually very
intuitive, since thermal fluctuations of the tangent vector along the backbone
of the polymer can be regarded as the time evolution of the quantum
rotor in the external field. Each polymer conformation hence represents
one particular time evolution of the quantum problem, in other words
`the path is the polymer'~\cite{Edwards78}.
The procedure for computing $\mathcal{Z}(\theta,\theta_i;s)$ we have outlined
above is thus formally equivalent to obtaining the spectral representation of a
quantum propagator in terms of the eigenstates of the associated Hamiltonian.

\section{Force-Extension in two dimensions}
\label{sec:forceext}
The eigenvalue equation, eqn.(\ref{sse}), can be transformed into
the canonical Mathieu differential equation by the change of variables
$\Theta = \pi/2-\theta/2$. As $\theta$ goes from $\pi$ to $-\pi$
about the direction of the force, $\Theta$ varies between $0$ and $\pi$. This
transformation yields,
\be
\frac{\partial^{2}Q}{\partial \Theta^{2}} + (p - 2q\cos 2\Theta)Q = 0 ,
\label{mde}
\ee
with $p=8E\kappa$ and $q=4f\kappa$. The solutions of this equation are the
Mathieu functions \cite{McLachlan47,Arscott64}. The Mathieu
differential equation always has a periodic solution and an aperiodic
solution~\cite{McLachlan47}. Since rotation of tangent vectors by $2\pi$ must
leave $Q$ unchanged, we desire a solution that is periodic in
$\theta$ with a period of $2\pi$, or, equivalently, periodic
in $\Theta$ with period $\pi$.

Solutions of eqn.(\ref{mde}) with a period of either $\pi$ or
$2\pi$ are called "basically periodic" in the Mathieu function
literature. A Mathieu function is basically periodic only for
specific values of the parameters  $p$ and $q$
~\cite{Abramowitz72,McLachlan47,Arscott64}. The value of $p$ that
makes the function basically periodic is called the
"characteristic value" of the Mathieu function, and is in fact a
continuous  function of
$q$~\cite{Abramowitz72,McLachlan47,Arscott64}. For a given $q$
there is a countable set of such characteristic-value functions, usually
denoted as $a_n(q)$ and $b_n(q)$, $n=0,1,2,\ldots$. The functions $a_n(q)$
are associated with the even Mathieu functions, $ce_{n}(a_n,q,\Theta)$, while
the functions $b_n(q)$ are associated with the odd Mathieu functions,
$se_{n}(b_n,q,\Theta)$; $ce$ and $se$ stand for
{\it cosine elliptic} and {\it sine elliptic}, respectively
\cite{Abramowitz72,McLachlan47,Arscott64}.
Furthermore, for $q\neq 0$, the Mathieu characteristic functions
satisfy,~\cite{Abramowitz72},
\be
a_{0}(q)< b_{1}(q) < a_{1}(q) < b_2(q) < a_2(q) < \ldots
\label{ladder}
\ee
We conclude that the periodicity of $Q(\Theta)$ imposes the quantization 
condition on the parameter $p=8E\kappa$, which must be from the countable set 
of Mathieu characteristic functions.

\begin{figure}
 \includegraphics[width=\columnwidth]{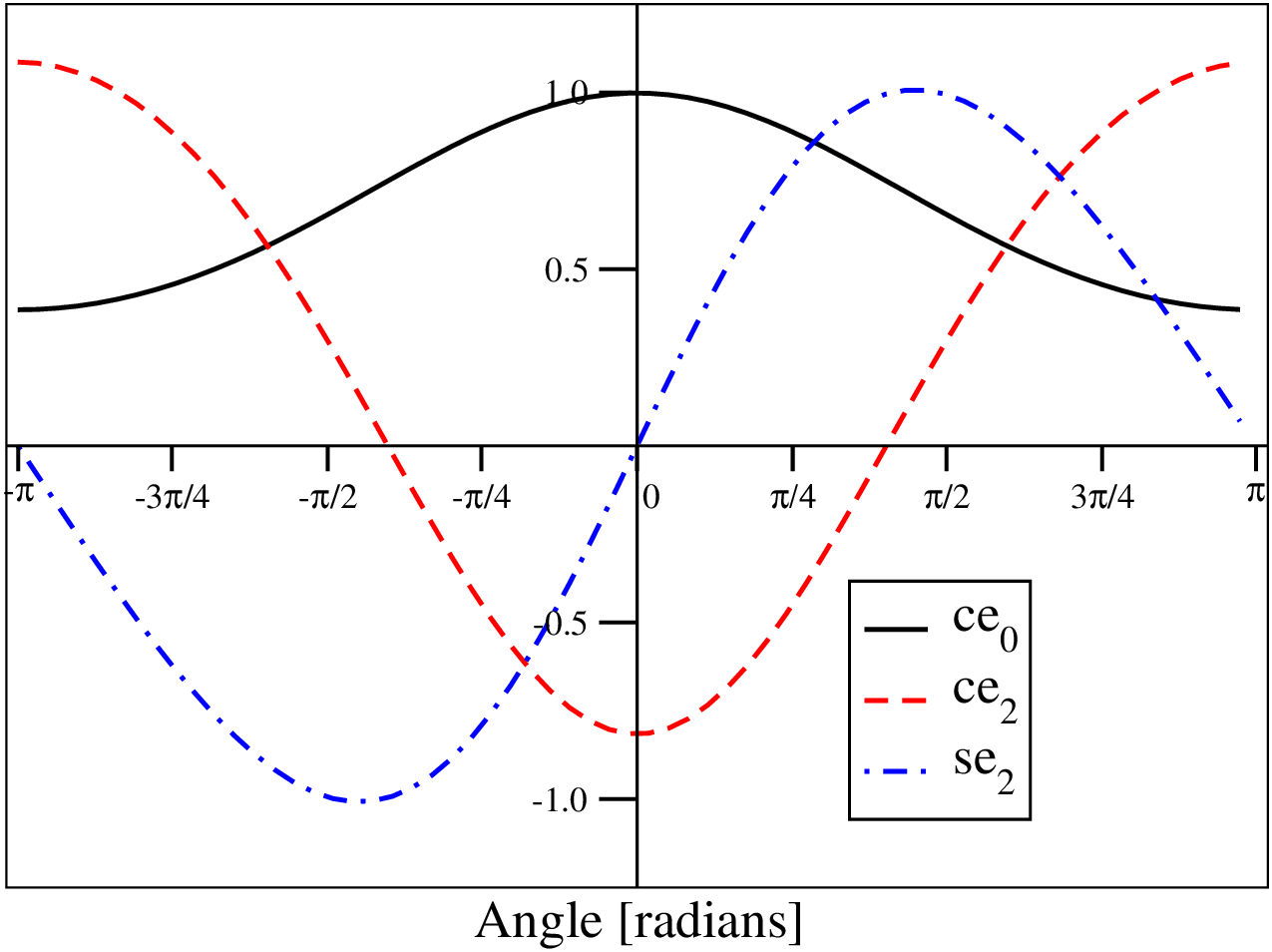}
 \caption{\label{wavfun} (Color online) Eigenfunctions $ce_0$, $se_2$ and $ce_2$
 plotted for $4f\kappa=1$ and for $\theta$ between $-\pi$ and $\pi$.}
 \end{figure}

\begin{figure}
 \includegraphics[width=\columnwidth]{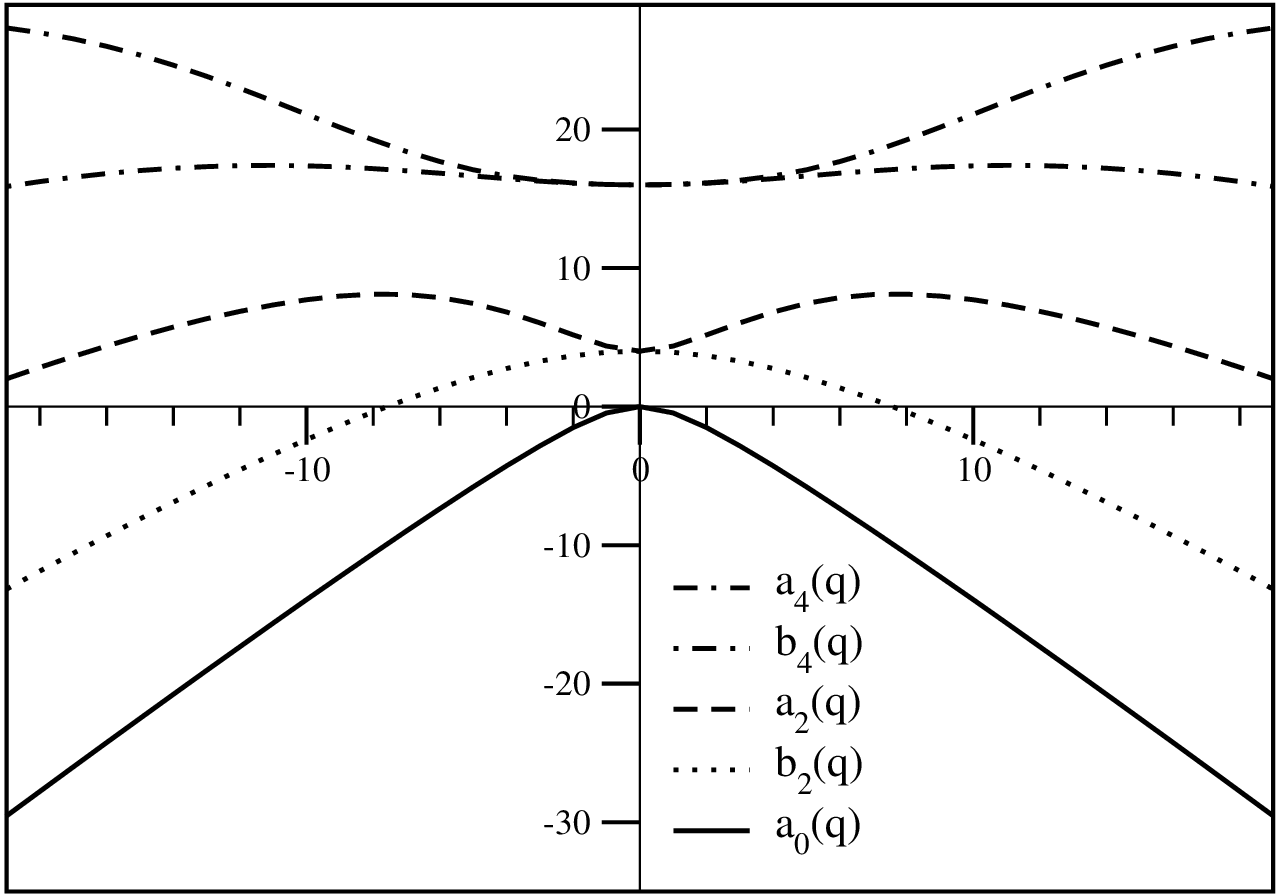}
 \caption{\label{chrfun} Characteristic functions $a_{0}(q)$, $b_{2}(q)$, 
$a_{2}(q)$, $b_{4}(q)$ and $a_{4}(q)$ plotted against $q$.}
 \end{figure}

The periodicity of the Mathieu functions for different values of $n$ can be
deduced from their expansions in sine and cosine series. There are
four types of basically periodic $ce$ and $se$ functions, and their expansions 
are
\bea
ce_{2n}(a_{2n},q,\Theta) &=& \sum_{r=0}^{\infty} A^{(2n)}_{2r}\cos 2r\Theta ,
\label{cosexp}\\
ce_{2n+1}(a_{2n+1},q,\Theta) &=& \sum_{r=0}^{\infty}A^{(2n+1)}_{2r+1} 
\cos (2r+1)\Theta ,\\
se_{2n+1}(b_{2n+1},q,\Theta) &=& \sum_{r=0}^{\infty}B^{(2n+1)}_{2r+1}
\sin(2r+1)\Theta ,\\
se_{2n+2}(b_{2n+2},q,\Theta) &=& \sum_{r=0}^{\infty}B^{(2n+2)}_{2r+2}
\sin(2r+2)\Theta .\label{sinexp}
\eea
Note that the coefficients $A$ and $B$ are functions of $q$. Perusal
of the above equations shows that
the class of functions that have period $\pi$ in $\Theta$, are
the $ce_{2n}$ and the $se_{2n+2}$ functions. The normalization convention
for the Mathieu functions is~\cite{McLachlan47}, \be
\int_{0}^{\pi}ce^{2}(a_n,q,\Theta)d\,\Theta =
\int_{0}^{\pi}se^{2}(b_n,q,\Theta)d\,\Theta = \frac{\pi}{2} .
\label{norm} \ee
The first few basically periodic Mathieu functions are plotted in 
fig.(\ref{wavfun}), while the
corresponding characteristic functions, are plotted in fig.(\ref{chrfun}).

With the eigenfunctions ($ce_{2n}$ and $se_{2n+2}$) and the corresponding 
eigenvalues ($a_{2n}$ and $b_{2n+2}$) of eqn.(\ref{sse}) in hand,
the spectral representation of the partition function of the 
worm-like chain is
\bea
\mathcal{Z}(\Theta_i,\Theta_f,L) &=&
\sum_{n=0}^{\infty} \left\{\frac{2}{\pi} ce_{2n}(a_{2n}, 4 f \kappa, \Theta_i)
  ce_{2n}(a_{2n}, 4 f \kappa, \Theta_f) \exp \left[-\frac{L}{8 \kappa}
  a_{2n}(4f\kappa)\right]\right.\nonumber \\
  &+& \left. \frac{2}{\pi} se_{2n+2}(b_{2n+2}, 4 f \kappa, \Theta_i)
  se_{2n+2}(b_{2n+2}, 4 f \kappa, \Theta_f)\exp \left[-\frac{L}{8\kappa}
  b_{2n+2}(4f\kappa)\right]\right\} , \label{prop}
  \eea
where $\Theta_i$ and $\Theta_f$ specify the tangent vectors at the two ends of
the chain. To obtain the partition function with no constraints on the initial
and final tangent vectors, we integrate over all
possible values of the initial and final angles. Using the
expansions, eqns.(\ref{cosexp}),(\ref{sinexp}) and~(\ref{norm}), we get
\be \mathcal{Z}(L) =
\sum_{n=0}^{\infty}2\pi \left(A_{0}^{(2n)}\right)^2
\exp \left[-\frac{L}{8\kappa} a_{2n}(4f\kappa)\right] .
\label{exactpart}
\ee
{}From the inequalities in eqn.(\ref{ladder}) it follows that, for $L$ large
compared to $\kappa$, the above expansions for $\mathcal{Z}$
are dominated by the term containing $a_{0}(4f\kappa)$ in the exponential. Even
for relatively short polymers, this term is a good
approximation to the full partition function. Namely, as can be seen from 
fig.(\ref{chrfun}), the difference between $b_2(4f\kappa)$ or $a_2(4f\kappa)$ 
and $a_0(4f\kappa)$ is $4$ at $f=0$ and increases in absolute value with 
increasing $f$. Therefore, even for 
polymer length $L\approx 8\kappa$, the second exponential in the expansion in 
eqn.(\ref{prop}) and eqn.(\ref{exactpart}) is at least $e^{-4}\approx 0.02$
times smaller than 
the first; the subsequent terms are exponentially smaller. 

It remains to consider the coefficients. In the case of eqn.(\ref{prop}) the
coefficients are
products of Mathieu functions and when $\theta_f \approx \theta_i \approx 0$,
which is a reasonable assumption for most experiments, the coefficient of the
first term is larger than any of the subsequent ones. 
In the case of eqn.(\ref{exactpart}), the coefficients $(A_0^{(2n)})^2$ are 
complicated functions of the dimensionless force $q$. However,  all of 
them satisfy $(A_0^{(2n)})^2<1$ and they are all of the same order of 
magnitude when $q>1$. When $q=0$, $A_0^{(0)}=1/\sqrt{2}$, and all other
coefficients are zero~\cite{McLachlan47}.
We conclude that  the approximation,
\be
\mathcal{Z} \approx c\exp[-\frac{L}{8\kappa}a_0(4f\kappa)] ,
\label{approxpart}
\ee
where $c$ is the appropriate coefficient
from either eqn.(\ref{prop}) or eqn.(\ref{exactpart}), depending on the boundary
conditions, is justified for polymer lengths $L>8\kappa$. 

We can now calculate the relative extension from eqn.(\ref{fext1}). The 
coefficient of the exponential yields a correction which is of order $\kappa/L$ 
and hence can be safely ignored in the long chain limit. The resulting force
extension relation is therefore given by,
\be \frac{\langle X\rangle}{L}=
-\frac{1}{8\kappa}\frac{da_{0}(4f\kappa)}{df} =
-\frac{1}{2}\frac{da_{0}(q)}{dq},  \label{fext} \ee
where $q=4f\kappa$, as before.

\begin{figure}
 \includegraphics[width=\columnwidth]{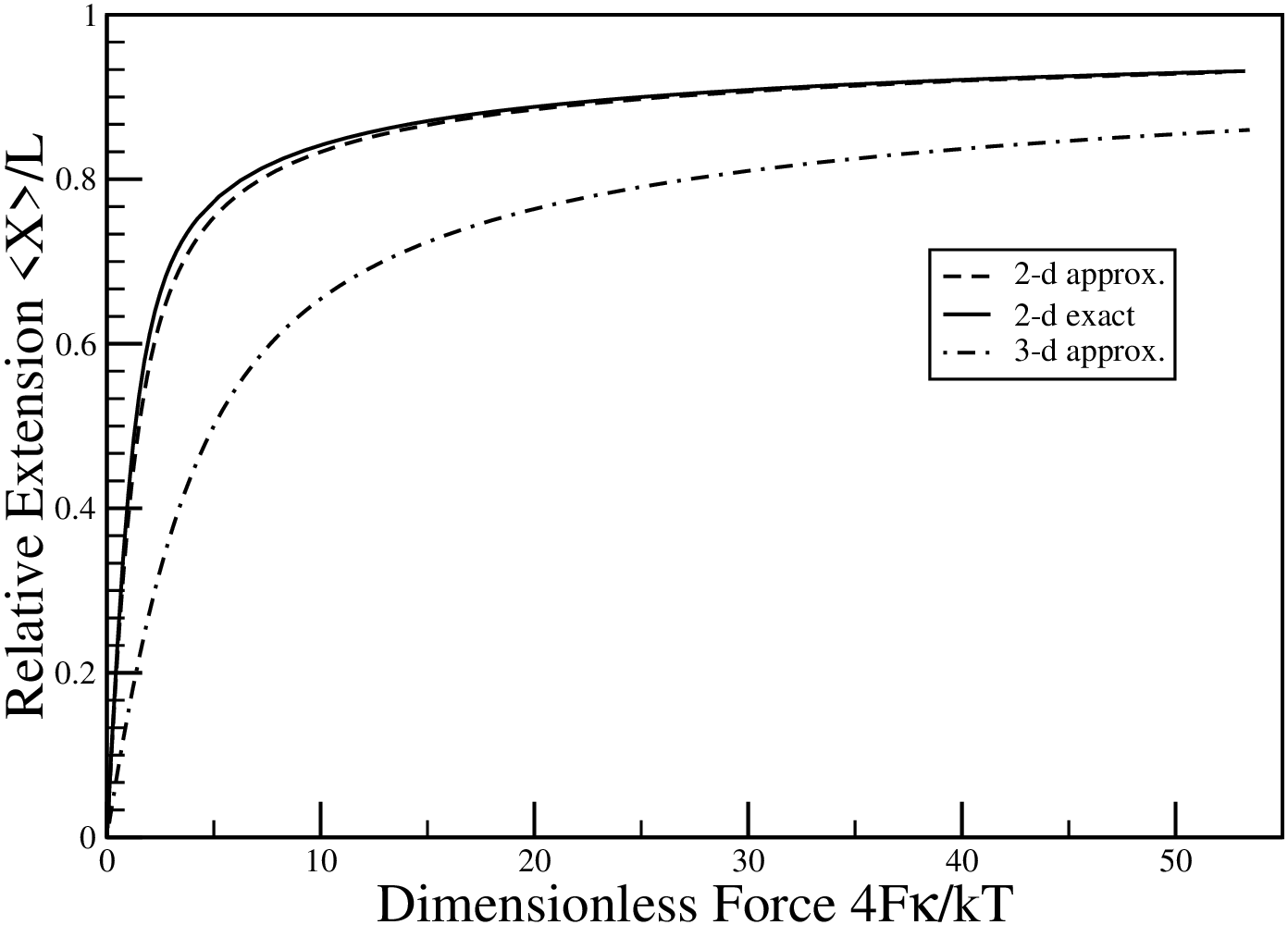}
 \caption{\label{2d3d} Force extension curves for semi-flexible polymers. The 
full line is the relative extension
 $\langle X\rangle/L$ plotted against  the dimensionless force, 
$q=4F\kappa/k_BT$ using the exact
 formula,  eqn.(\ref{fext}). The dashed line is the interpolation
 formula,  eqn.(\ref{interpol}). For comparison, the force-extension
 curve in three-dimensions is plotted as a dot-dashed line, using the 
 approximate interpolation formula described in Ref.~\cite{Marko95c}.}
 \end{figure}

The Mathieu Characteristic functions are interesting in themselves.
So far no exact expression for them is
known in terms of other functions, though good polynomial
approximations are available for small and large $q$. They
are usually expressed as roots of an equation involving infinite
continued fractions ~\cite{Abramowitz72, McLachlan47,Arscott64}.
They can also be shown to be the roots of an infinite tri-diagonal
determinant equation~\cite{Arscott64}. Both these
representations allow numerical calculation as well as numerical
differentiation by mathematical programs like
Mathematica~\cite{Wolfram03}. Mathematica executes the desired
function as \texttt{MathieuCharacteristicA[r,q]}, which we use to
generate the plot of the relative extension as a function of the
applied force, shown in fig.(\ref{2d3d}).

It is worthwhile to note that the higher order moments of the probability
distribution of the end-to-end distance can
be obtained by successive differentiation of the partition 
function,
\be
\langle X^n \rangle =
\frac{1}{\mathcal{Z}}\frac{\partial^n\mathcal{Z}}{\partial F^n} .
\ee

Having computed the exact expression for the force-extension relation for a
two-dimensional worm-like chain in terms of Mathieu functions, in the
next section we consider the small and large force limits of 
eqn.(\ref{fext}), which are readily obtained using more elementary 
methods.

\section{Limits of small and large force}
\label{sec:limits}
In order to gain a better understanding of the derived force-extension relation, eqn.(\ref{fext}),
we analyze it in the small force ($q = 4 f \kappa \ll 1$) and the large force ($q \gg 1$) limits.

{}For small forces, we use the power series expansion~\cite{Abramowitz72} of $a_{0}(q)$,
\be
a_{0}(q)=-\frac{q^{2}}{2} + \frac{7q^{4}}{128} +
O(q^{6}) .
\ee
Therefore, to first order in $q$, the extension becomes
\be
\frac{\langle X \rangle}{L}= \frac{1}{2}q = 2 f\kappa .
\label{lowforce}
\ee

It is reassuring to find that a perturbation analysis of eqn.(\ref{sse}), in 
the limit of low force, yields the same result.
If the potential $f\cos \theta$ is treated as a perturbation, the unperturbed
equation describes a free particle on a ring,
\be
\frac{1}{2\kappa}\frac{d^2 Q_n}{d\theta^2} = -E^{(0)}_n Q_n . \label{lowfse}
\ee
The solutions of this equation are sine and cosine functions, except for the $E^{(0)}_0=0$
solution which is $Q_0 = 1/\sqrt{2\pi}$. The eigenvalues are given by,
\be
E_n^{(0)} = \frac{n^2}{2 \kappa}, \quad n=0,1,2, ....
\label{lowfev}
\ee
The first order correction to the zero-eigenvalue is zero.
Using orthogonality properties of sine and cosine functions, we obtain the 
second order correction
\be
E_0^{(2)} = -\kappa f^2 .
\ee
Using $\langle X \rangle/L = - dE_0/df$, which follows from eqn.(\ref{fext1}), 
and the fact that in the long-chain limit ${\mathcal Z}\sim \exp(-E_0 L)$,
we get eqn.(\ref{lowforce}).

Let us now consider the limit of large forces. We use the asymptotic
expansions given in Ref.~\cite{Abramowitz72},
\be a_{0}(q) \sim -2q +
2q^{\frac{1}{2}} - \frac{1}{4} - \frac{1}{32}q^{-\frac{1}{2}} +
O(q^{-1}) .
\ee
This gives us, after ignoring terms of order $q^{-1/2}$ and smaller,
\be
\frac{\langle X \rangle}{L} = 1-\frac{1}{2\sqrt{q}} = 1 -
\frac{1}{4\sqrt{f\kappa}} .
\label{hf}
\ee

This formula can be derived independently by noting that at high forces the
polymer is highly extended in the direction of the force, with only small
transverse fluctuations, hence the angle $\theta$ that $\hat{\bm{t}}$ makes
with $\hat{\bm{x}}$ is very small~\cite{Marko95c}. We can therefore use the 
approximation
\be
\cos \theta \approx 1 - \frac{1}{2}\theta^2 .
\ee
This transforms the differential equation to be solved, eqn.(\ref{sse}), into
a form that is equivalent to the Schr\"{o}dinger equation for a
one-dimensional simple harmonic oscillator with $\hbar = 1$ and
angular frequency $\omega = \sqrt{f/\kappa}$,
\be
\frac{1}{2 \kappa}\frac{\partial^2 Q_n}{\partial \theta^2} -
\frac{1}{2}f\theta^2 Q_n = -(E_n+f)Q_n . \label{sho}
\ee
The eigenfunctions $Q_n$ are gaussian functions multiplied by Hermite
polynomials, and the eigenvalues are given by the well known expression,
\be
E_n+f = \left(n + \frac{1}{2}\right)\sqrt{f/\kappa} \ .
\ee
Therefore the partition function can be written as,
\be
\mathcal{Z}(\theta_f,\theta_i;L) = \sum_{n=0}^{\infty}
Q_n(\theta_i)Q_{n}(\theta_f)\exp \left[- E_n L\right] .
\ee
Taking only the first term, in the limit of large-$L$, and integrating over all
values of the initial and final angles, we get,
\be
\mathcal{Z} \sim \exp[- E_0 L] =
\exp \left[ \left(f - \frac{1}{2}\sqrt{f/\kappa} \right)L \right] \ .
\ee
Using this result in  eqn.(\ref{fext1}) immediately gives  eqn.(\ref{hf}), as
expected.

In fig.(\ref{2d3d}) we have plotted a comparison of the force-extension
curves from eqn.\ (\ref{fext}) and the interpolating formula of
Ref.~\cite{Marko95c}. The $2d$ curve is found to lie strictly above the $3d$
curve at non-zero force. This is expected by entropic arguments since a three
dimensional polymer has larger entropy, hence a larger force is needed
to extend it by the same amount.
Note that in two dimensions DNA is highly stretched even at relatively low
forces.

It is also possible to write down an approximate formula for the force-extension
of a two-dimensional polymer, which tends to the exact force-extension relation
in the two limits, and  is approximately true to within less than 10\% in the
remaining  regime (See fig.(\ref{2d3d})),
\be
16f\kappa = 6 \frac{\langle X \rangle}{L} - 1  +
\frac{1}{\left(1- \frac{\langle X \rangle}{L}\right)^2} .
\label{interpol}
\ee

\section{Semiflexible polymer in a nematic field}
\label{sec:nematic}

Another way to stretch a polymer is to put it in an aligning field, such as
that produced by a nematic liquid crystal. In a recent experiment,
semi-flexible polymers were stretched by dissolving them in the nematic phase
of rod-like {\it fd}-viruses \cite{Dogic04}. The nematic potential experienced
by the polymer can be written as \cite{Warner85a,Spakowitz04a},
\be
\frac{V}{k_BT}=-\Gamma \int_0^L d\,s \left\{(\hat{\bm{t}}\cdot
\hat{\bm{n}})^2 - 1/2 \right\} , \label{nempot}
\ee
where the square arises due to the reflection invariance of the unit nematic
director $\hat{\bm{n}}$, which we assume again is in the
$\hat{\bm{x}}$-direction. The coupling $\Gamma$, between the nematic field
and the polymer depends upon the order parameter of the nematic liquid crystal
in which the polymer is embedded. When the liquid crystal undergoes a phase
transition from the isotropic to the nematic phase, $\Gamma$ increases and the
polymer stretches.
This mean-field form of the interaction ignores the fluctuations in the
nematic order parameter ~\cite{Kamien92, DeGennes82}.

With the help of the trigonometric identity, $\cos^2 \theta -1/2= (\cos
 2\theta)/2$, the
potential in eqn.(\ref{nempot}) becomes similar to that of the
constant pulling-force case, but with $\cos \theta $ replaced by
$\cos 2\theta$. Following the treatment detailed in a previous section, 
the partition function can again be
expressed in a spectral representation, using the characteristic
functions of the Mathieu differential equation, which for this case is,
\be
\frac{\partial^2 Q}{\partial
\theta^2}+ \Gamma \kappa Q \cos 2\theta = -2E\kappa Q .
\label{nemse}
\ee
As before the $ce_{2n}$ and $se_{2n+2}$ functions are the appropriate
eigenfunctions. In the large-$L$ limit, the smallest eigenvalue again
dominates  the spectral expansion and the partition function is well
approximated by,
\be
\mathcal{Z} = 2 \pi(A_0^{(0)})^2 \exp \left[-\frac{L}{2\kappa}
a_{0}\left(\frac{\Gamma \kappa}{2}\right)\right] .
\label{nempart}
\ee
The derivative of the free energy with respect to the coupling
$\Gamma$ is now
\be
S\equiv \frac{\partial \ln\mathcal{Z}}{\partial \Gamma} =
\langle \int_0^L d\,s \{\cos^2\theta - 1/2\}\rangle ,\label{nems}
\ee
which is a measure
of the degree of alignment of the polymer with the external field. It can be
seen that when the polymer lies perfectly along the $\hat{\bm{x}}$ direction,
then $S=L/2$, but when the polymer conformation is
completely random, $S=0$, since $\langle cos^2\theta \rangle = 1/2$. 
Therefore $2S/L$ is the nematic order parameter of the polymer. The 
partition function eqn.(\ref{nempart})
enables us to calculate $S/L$ as a function of the coupling strength $\Gamma$,
\be
\frac{S}{L}= -\frac{1}{4}\frac{da_0(q)}{d q} ,
\label{nemord}
\ee
where now $q=\Gamma \kappa/2$, and we have ignored terms of order $\kappa/L$.
Using the above result we can also obtain the
approximate relative extension of the polymer in the direction of the nematic
director at high coupling $\Gamma$. Due to the reflection invariance of the
nematic director, the extension along that direction is defined as,
\be
\langle \mid X \mid \rangle =
\langle \mid \int_0^L d\,s \cos \theta \mid \rangle . \label{nemextdef}
\ee
When the coupling between the polymer and the field is large the polymer
conformation would consist of roughly straight sections which have small
fluctuations in the $\hat{\bm{y}}$ direction, interspersed by hairpin bends.
A simple thermodynamic argument, based on comparison of the energy cost of a
hairpin with the entropy of a hairpin, leads to the conclusion that the 
typical length of the straight segments grows as the exponential of the square 
root of the coupling strength,
\be
L_{nem} \sim \pi\sqrt{\pi \kappa\Gamma^{-1}} 
\exp\left(2\sqrt{\pi \kappa \Gamma}\right) .
\ee
This is similar to the relation discussed in ref.~\cite{DeGennes82}. 
For DNA, $\kappa \approx 50 nm $, therefore $L_{nem}\approx 50 \mu m$ when 
$\Gamma = 0.05 nm^{-1}$, and $L_{nem}\approx 3 m$ when $\Gamma$ is an order of 
magnitude larger. If the length of the polymer is smaller than $L_{nem}$, 
hairpin bends are ruled out, and $\cos \theta$ does not change
sign along the polymer contour. Without loss of generality, we can take it to
be positive and remove the absolute value in eqn.(\ref{nemextdef}). 
This allows us to write the following approximate formula for the
extension,
\be
\langle X\rangle \approx
\langle\int_{0}^{L}d\,s \sqrt{1 - \sin^2 \theta}\rangle
\approx \langle\int_{0}^{L}d\,s
\left\{1-\frac{1}{2}\sin^2\theta\right\}\rangle . \label{nemextapp}
\ee
By arguments similar to those made earlier for the polymer under a constant 
force,
we know that eqn.(\ref{nempart}) is a good approximation for the partition
function when $L/\kappa>>1$, and we may use it to calculate the right-hand 
side of eqn.(\ref{nemextapp}). 
The relative extension of the polymer along $\hat{\bm{x}}$ when $L < L_{nem}$ 
and $L/\kappa>>1$ can therefore be expressed as,
\be
\frac{\langle X \rangle}{L} \approx \frac{3}{4} -
\frac{1}{8}\frac{d a_0(q)}{d q} , \label{nemext}
\ee
where $q = \Gamma \kappa /2$. In the limit $q>>1$, the relative extension
has the simple form, $1-1/8\sqrt{q}$. These results can be used to estimate 
$\Gamma$ from experiments that measure the relative extension.

It should be noted that the exponential dependence of $L_{nem}$ on the square 
root of $\Gamma$ implies that when $\Gamma\kappa>1$, $L_{nem}$ could be quite
large, hence it would be easy to satisfy both $L<L_{nem}$ and $L/\kappa>>1$. 
For example it was estimated above that $L_{nem}$ was as much as 
$50 \mu m$ for DNA when $\Gamma \kappa \approx 2.5$. 
At this value of $\Gamma$, eqn.(\ref{nemext}) predicts that the molecule 
would have extended to $87\%$ of its contour length. While to our knowledge,
experimental estimates of $\Gamma$ for DNA embedded in a nematic liquid 
crystal do not exist at present, the experiment of ref.~\cite{Dogic04} 
measured values up to $\Gamma \approx 800 (\mu m)^{-1}$ for 
wormlike micelles ($\kappa \approx 0.5 \mu m$) in a solution of fd-viruses. 
The regime of validity of eqn.(\ref{nemext}) is therefore expected to be 
quite large. 
These arguments also imply that a dramatic straightening of the polymer 
should take place as the liquid crystal goes through an isotropic-nematic phase 
transition. This indicates that there is considerable scope for using liquid 
crystals for stretching DNA in the laboratory.

\section{Conclusion}
At present several schemes exist to calculate the partition function
of the worm-like chain  models in both two and three dimensions, to any 
degree of
accuracy~\cite{Marko95c,Samuel02a,Stepanow04, Spakowitz04a}. With the exception
of ref.~\cite{Spakowitz04a}, all use eigenfunction expansions to obtain
the partition function numerically. Ref.~\cite{Spakowitz04a} uses a
novel method to express the partition function in terms of infinite continued
fractions.

This multitude of approaches only underlines the rich physics of
the worm-like chain model, and also hints at interesting mathematical
connections. In the present work, the use of
Mathieu functions to derive an exact partition function yields a
simple closed-form expression for the free energy of the worm-like chain 
in the long chain limit. This allowed us to calculate the force-extension
relation in two dimensions in terms of Mathieu characteristic 
functions. We also discussed the application of this result to the interaction 
between a semi-flexible polymer and a nematic field, and derived the nematic 
order parameter and average extension of the polymer in a strong field. It 
should be noted that self-avoidance, which is more important in two dimensions
than in three, has been ignored in our analysis. We  address
this question, as well as stretching by a force that does not remain constant
along the contour, such as that produced by an electric field, in a future 
publication ~\cite{Horitemp}.

This work was supported by the NSF through grants DMR-9984471 and DMR-0403997.
JK is a Cottrell Scholar of Research Corporation.

\end{document}